\documentclass[journal,12pt,onecolumn,draftclsnofoot,]{IEEEtran}

\usepackage[cmex10]{amsmath}
\usepackage[utf8]{inputenc}
\usepackage{multirow}
\usepackage{soul}
\usepackage{algorithm}
\usepackage{algpseudocode}
\usepackage{graphicx}
\usepackage{dsfont}
\usepackage{xcolor}
\usepackage{cite}

\newcommand{\probP}{\text{I\kern-0.15em P}}

\usepackage[utf8]{inputenc}
\usepackage{mathtools}
\usepackage[mathscr]{euscript}
\usepackage{amsmath}
\usepackage{amssymb}
\usepackage{amsfonts}
\usepackage{amssymb}
\usepackage{amsmath}
\usepackage{verbatim}
\usepackage[T1]{fontenc}
\usepackage[utf8]{inputenc}
\usepackage{latexsym}
\usepackage{enumerate}
\usepackage{indentfirst}
\usepackage{calligra}
\usepackage{ulem}
\usepackage{graphicx}
\usepackage{array}
\usepackage{float}
\usepackage{bbm}
\usepackage{color, colortbl}

\definecolor{LightCyan}{rgb}{0.88,0.75,1}	
\definecolor{Gray}{gray}{0.9}
\usepackage{geometry}

\usepackage{mleftright}
\usepackage{colortbl}
\makeatletter
\let\old@ps@headings\ps@headings
\let\old@ps@IEEEtitlepagestyle\ps@IEEEtitlepagestyle
\def\psccfooter#1{%
 \def\ps@headings{%
 \old@ps@headings%
 \def\@oddfoot{\strut\hfill#1\hfill\strut}%
 \def\@evenfoot{\strut\hfill#1\hfill\strut}%
 }%
 \def\ps@IEEEtitlepagestyle{%
 \old@ps@IEEEtitlepagestyle%
 \def\@oddfoot{\strut\hfill#1\hfill\strut}%
 \def\@evenfoot{\strut\hfill#1\hfill\strut}%
 }%
 \ps@headings%
}
\makeatother

\usepackage{soul}
\usepackage{multirow}
\usepackage{soul,color}
\usepackage{graphicx}
\usepackage{dsfont}
\usepackage{subcaption}

\usepackage[utf8]{inputenc}
\usepackage{mathtools}
\usepackage[mathscr]{euscript}
\usepackage{amsmath}
\usepackage{amssymb}
\usepackage{amsfonts}
\usepackage{verbatim}
\usepackage[T1]{fontenc}
\usepackage[utf8]{inputenc}

\usepackage{latexsym}
\usepackage{enumerate}
\usepackage{indentfirst}
\usepackage{calligra}
\usepackage{ulem}
\usepackage{multirow}

\usepackage{array}
\usepackage{float}
\usepackage{bbm}

\usepackage{eurosym}

\usepackage{hyperref}

\usepackage{caption}
\usepackage{tikz}
\usepackage{pgfplots}

\pgfplotsset{compat=1.8}
\usepgfplotslibrary{statistics}
\pgfmathdeclarefunction{fpumod}{2}{%
 \pgfmathfloatdivide{#1}{#2}%
 \pgfmathfloatint{\pgfmathresult}%
 \pgfmathfloatmultiply{\pgfmathresult}{#2}%
 \pgfmathfloatsubtract{#1}{\pgfmathresult}%
 \pgfmathfloatifapproxequalrel{\pgfmathresult}{#2}{\def\pgfmathresult{5}}{}%
 }

\usepackage{tabularx}
\usepackage{flushend}
\usepackage{textcomp}
\usepackage{xcolor}
\usepackage{import}

\usepackage{csvsimple}
\usepackage{tabularx}
\usepackage{adjustbox}

\usetikzlibrary{trees,decorations,shadows}
\tikzset{level 1/.style={sibling angle=45,level distance=4mm}}
\usetikzlibrary{arrows.meta}
\usepackage{forest}
\usetikzlibrary{external}

\let\oldtikzexternalgetnextfilename\tikzexternalgetnextfilename \renewcommand{\tikzexternalgetnextfilename}[1]{\oldtikzexternalgetnextfilename{#1}\expandafter\tikzsetnextfilename\expandafter{#1}}

\usepackage{pgfplotstable}
\usepackage[outline]{contour}
\contourlength{1.2pt}

\pgfplotsset{compat=1.13} 

\usetikzlibrary{spy}
\usetikzlibrary{calc}
\usetikzlibrary{fadings}
\usetikzlibrary{patterns}
\usetikzlibrary{shadows}
\usetikzlibrary{mindmap}
\usetikzlibrary{backgrounds}
\usetikzlibrary{shapes.symbols}
\usetikzlibrary{shapes.multipart}
\usetikzlibrary{shapes.geometric}
\usetikzlibrary{automata,positioning}
\usetikzlibrary{decorations.fractals} 
\usetikzlibrary{decorations.markings}
\usetikzlibrary{decorations.pathreplacing}
\usetikzlibrary{decorations.pathmorphing}
\usepackage{svg}
 
\usepackage{bm}

\usepackage{nomencl}
\makenomenclature
\tikzset{edge from parent/.style={segment angle=10,draw}}

\tikzset{
 my rounded corners/.append style={rounded corners=2pt},
}

\def\BibTeX{{\rm B\kern-.05em{\sc i\kern-.025em b}\kern-.08em
 T\kern-.1667em\lower.7ex\hbox{E}\kern-.125emX}}

\renewcommand{\nomgroup}[1]{%
 \ifthenelse{\equal{#1}{O}}{\item[\textit{Operators}]}{%
 \ifthenelse{\equal{#1}{I}}{\item[\textit{Indices}]}{%
 \ifthenelse{\equal{#1}{A}}{\item[\textit{Acronyms}]}{%
 `\ifthenelse{\equal{#1}{V}}{\item[\textit{Variables and parameters}]}{}}}}}
\usepackage{scalerel}
\usetikzlibrary{svg.path}
\definecolor{orcidlogocol}{HTML}{A6CE39}
\tikzset{
 orcidlogo/.pic={
 \fill[orcidlogocol] svg{M256,128c0,70.7-57.3,128-128,128C57.3,256,0,198.7,0,128C0,57.3,57.3,0,128,0C198.7,0,256,57.3,256,128z};
 \fill[white] svg{M86.3,186.2H70.9V79.1h15.4v48.4V186.2z}
 svg{M108.9,79.1h41.6c39.6,0,57,28.3,57,53.6c0,27.5-21.5,53.6-56.8,53.6h-41.8V79.1z M124.3,172.4h24.5c34.9,0,42.9-26.5,42.9-39.7c0-21.5-13.7-39.7-43.7-39.7h-23.7V172.4z}
 svg{M88.7,56.8c0,5.5-4.5,10.1-10.1,10.1c-5.6,0-10.1-4.6-10.1-10.1c0-5.6,4.5-10.1,10.1-10.1C84.2,46.7,88.7,51.3,88.7,56.8z};
 }
}

\newcommand\orcidicon[1]{\href{https://orcid.org/#1}{\mbox{\scalerel*{ \begin{tikzpicture}[yscale=-1,transform shape]
 \pic{orcidlogo};
 \end{tikzpicture}
 }{|}}}}

\begin{document}
%
\title{\huge{Analyzing electric vehicle, load and photovoltaic generation uncertainty using publicly available datasets}}

\author{Md~Umar~Hashmi$^{1}$,~\IEEEmembership{Senior~Member,~IEEE}~\orcidicon{0000-0002-0193-6703},~Domenico~Gioffrè$^{2}$,~\IEEEmembership{Member~IEEE}~\orcidicon{0000-0002-5632-7892},~Simon~Nagels$^{1}$,~\IEEEmembership{Member,~IEEE}~\orcidicon{0009-0008-6487-6526}, and~Dirk~Van~Hertem$^{1}$,~\IEEEmembership{Senior~Member,~IEEE}~\orcidicon{0000-0001-5461-8891}
\thanks{Corresponding author email: mdumar.hashmi@kuleuven.be}
\thanks{$^{1}$M.U.H., S.N. and D.V.H. are with KU Leuven, division Electa in Leuven \& EnergyVille in Genk, Belgium}
\thanks{$^{2}$D.G. is with the Department of Energy, Politecnico di Milano, Milan, Italy}
} 

 

 
 



\maketitle

\begin{abstract}
This paper aims to analyze three publicly available datasets for quantifying seasonal and annual uncertainty for efficient scenario creation. The datasets from Elaad, Elia and Fluvius are utilized to statistically analyze electric vehicle charging, normalized solar generation and low-voltage consumer load profiles, respectively. 
Frameworks for scenario generation are also provided for these datasets.
The datasets for load profiles and solar generation analyzed are for the year 2022, thus embedding seasonal information.
An online repository is created for the wider applicability of this work. Finally, the extreme load week(s) are identified and linked to the weather data measured at EnergyVille in Belgium.
\end{abstract}


{\textbf{Disclaimer}: This paper is a preprint of a paper submitted to and presented at the IEEE International Conference on Power System Technology (PowerCon) 2024 Kathmandu, Nepal. 
The final version of the paper will be available at IEEE xplore.}

\pagebreak

\tableofcontents

\pagebreak

\section{Introduction}
The lack of quality datasets often constrains the pursuit of optimal decisions in power networks, especially distribution networks (DN). 
This is due to insufficient measurements and the scarcity of representative historical datasets. This work aims to analyze three publicly available datasets in Belgium and the Netherlands. We quantify the seasonal and annual parameter uncertainties that can be used to generate meaningful scenarios. These scenarios can be utilized for reliable operational and planning decision-support activities for the DNs.

The first dataset analyzed is from \href{https://elaad.nl/en/}{ElaadNL}. Elaad is an initiative of the Dutch grid operators to make the power grid future ready for accommodating large-scale electric vehicle (EV) penetration while also facilitating smart charging \cite{ElaadData}. The Elaad dataset comprises of more than 10,000 charging session details for charging power ranging from 0 to 23 kW, with maximum charged energy in a session exceeding 80 kWh.
We propose a scenario generation framework based on the stochasticity of charging power, arrival, departure, connection and charge time. 

The second dataset analyzed is from \href{https://www.elia.be/en/}{Elia}, a transmission system operator in Belgium. Elia provides a historical data dashboard. In this work, we use Elia's extensive data platform for quantifying solar generation uncertainty \cite{solarELIA}. We calculate seasonal and annual quartiles to identify realistic normalized photovoltaic (PV) generation scenarios.
The total PV generation is a function of installed capacity and the normalized PV generation. The former uncertainty is modelled via a separate random variable, thus decoupling the sources of uncertainty, as in \cite{koirala2022decoupled}.

The final dataset analyzed is from \href{https://www.fluvius.be/nl}{Fluvius}, a distribution system operator (DSO) in Flanders, Belgium.
The Fluvius dataset comprises 1300 consumer load profiles for 1-year \cite{FluviusData}. This high-quality dataset is well-labelled, and appropriate measures have been taken to eliminate missing data.
We identify the maximum probability of load peaks and reverse power peak occurrences for this dataset. Finally, we link these representative days (as in \cite{belderbos2015accounting}) with weather parameter measurements at \href{https://energyville.be/en/}{EnergyVille} in Belgium. The weather measurements include ambient temperature, wind speed, humidity, wind direction, Global Horizontal Irradiance (GHI), diffuse horizontal irradiance (DHI) and rainfall. 
\vspace{-5pt}
\vspace{-5pt}
\subsection{Related Literature \& Use Cases}
Scenario generation using historical load profile data is explored in \cite{soenen2023scenario, chen2018model, dong2022data, ma2013scenario, tang2018efficient}.
In this subsection, we detail the previous use cases that utilized the datasets explored in this paper.


\subsubsection{Elaad dataset usecase}
Klaassen et al. \cite{klaassen2022bottom} adopted a bottom-up load modelling strategy to assess the impact of EVs on DNs. The dataset randomly samples load profiles of individual charging sessions, which are then distributed to charging points. Kern et al. \cite{kern2023detection} developed a hybrid intrusion detection system method consisting of regression-based charging session forecasting and anomaly detection to prevent potential security incidents. Mangipinto et al. \cite{mangipinto2022impact} used the dataset to validate their developed open-source model, which simulates aggregate mobility and charging patterns for large electric vehicle fleets at high (1-min) temporal resolution against empirical data. Akil et al. \cite{akil2021soc} utilized the dataset for a dynamic charge coordination method based on the EVs' battery state of charge. This method has been compared to the uncoordinated charging method for integrating PV and storage systems with load balancing for EVs.

\subsubsection{Fluvius dataset usecase}
Soenen et al. \cite{soenen2023scenario} compared a machine-learning (ML) sampling method and an expert-based sampling method to generate residential load profile scenarios for new unsupervised customers with known basic attributes (e.g. yearly consumption, total PV capacity, etc.) from historical load profiles provided by Fluvius. It's concluded that both methods outperformed random sampling and standard load profiles, while the ML method performed better. Similarly, Claeys et al. \cite{claeys2023stochastic} proposed a wavelet-decomposition-based method to generate residential load profiles with realistic variability and include peak values based on historical load profiles of Fluvius. The strengths and limitations of the generated load profiles were discussed using load profile analysis and an EV hosting capacity case study.


\subsubsection{Elia solar generation}
Authors in \cite{heymans2023modelling} use the Elia load profile dataset to compare the accuracy of various forecasting methods to predict the Belgian electricity load profile. The proposed generalized additive model 
includes daily, weekly, and seasonal load profile variability. 
The impact of intermittency of renewable energy sources on the power system is reviewed in \cite{asiaban2021wind}. This paper further analyses the variability of these energy sources discusses forecasting methods, and outlines how demand response and energy storage can accommodate this variability.


\subsection{Contributions of the paper}
The contributions of the paper are as follows:\\
$\bullet$ \textit{Analyzing publicly available datasets}: statistically analyze datasets for electricity consumption by LV consumers, electric vehicle charging profiles and solar generation profiles quantified in annual and seasonal parameter variations. Subsequently, scenarios can be generated by uncertainty quantification for 1 year of data, that implicitly embeds seasonality. \\
$\bullet$ \textit{Repository for efficient scenario generation}: an online repository is created to quantify seasonality, parameter variation as probability density functions (PDFs), and conditional probability assessment for efficient scenario generation for PV and EV profiles. Further, consumer load meta attributes are identified that can be used to select appropriate load profiles to generate realistic scenarios for operational and planning studies for DNs.


This paper is organized as follows:
Section \ref{EVdataset} details the Elaad dataset analysis along with a scenario generation framework.
Section \ref{section3} quantifies the Elia dataset's seasonal and annual solar generation uncertainty.
Section \ref{section4} describes the analysis for Fluvius dataset for 1300 consumers. The metadata of all these consumers is extracted.
Section \ref{section5} concludes the paper.

\pagebreak

\section{Elaad EV charging dataset} \label{EVdataset}
This section analyzes the open dataset provided by Elaad \cite{ElaadData}, comprising an overview of 10,000 random charging events from 2019 at public charging stations operated by Elaad. The objective is to identify statistical correlations among the available charging variables and utilize them to generate realistic scenarios of EV charging sessions.

\begin{figure}[!htbp]
    \centering
    \includegraphics[width=0.9\textwidth]{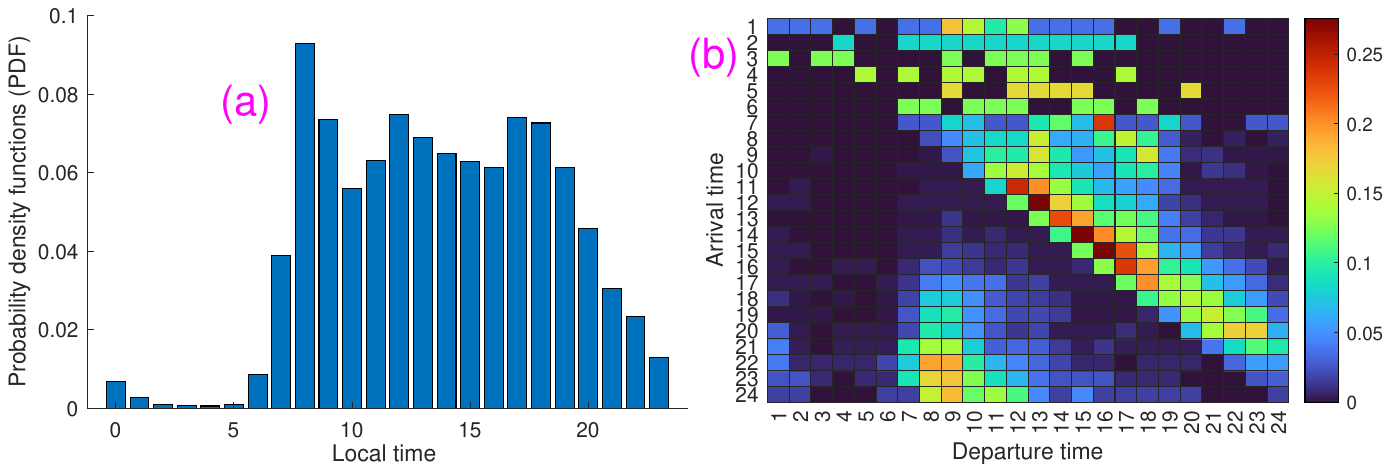}
    \vspace{-4pt}
    \caption{\small{Probability distribution \(t_{arr}\) (a) and conditional probability distribution \(t_{dep}\) | \(t_{arr}\) (b).}}
    \label{fig:evpdfs}
\end{figure}

\begin{figure}[!htbp]
    \centering
    \includegraphics[width=0.9\textwidth]{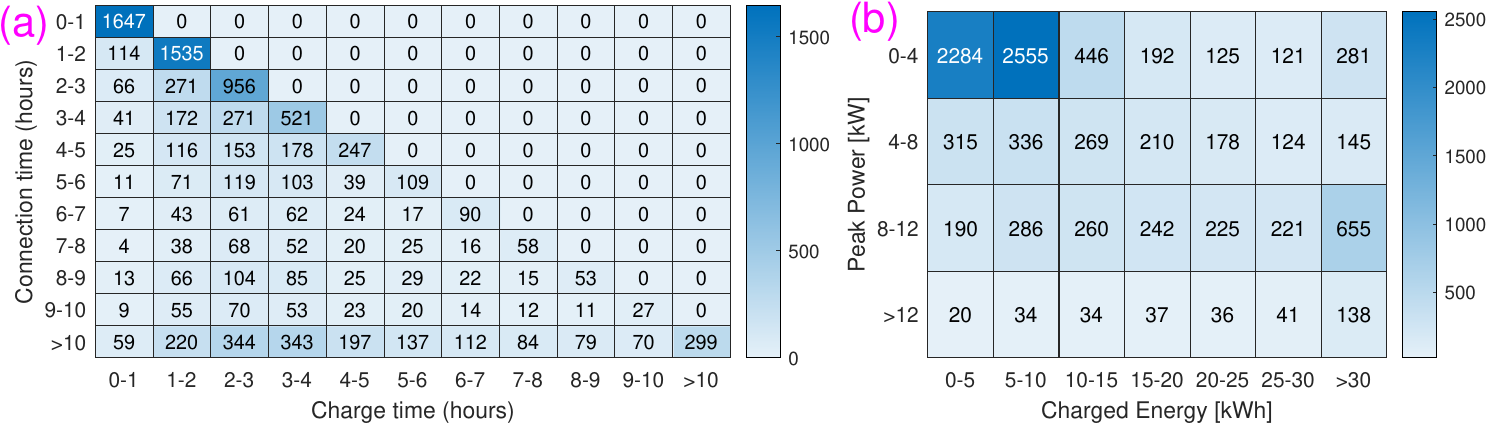}
    \vspace{-4pt}
    \caption{\small{(a) shows the correlation between charge time and connected time and (b) charged energy and peak power.}}
    \label{fig:evoccur}
\end{figure}

In the dataset, each charging session is characterized by arrival time (\(t_{arr}\)), departure time (\(t_{dep}\)), connection time (\(\Delta t_{conn}\)), charge time (\(\Delta t_{ch}\)), peak power (\(P_{peak}\)) and charged energy (\(EE_{ch}\)). Fig. \ref{fig:evpdfs}a illustrates the probability distribution of arrival times at the public charging point, indicating a widespread occurrence during the morning and evening and reduced activity at night. Fig. \ref{fig:evpdfs}b displays the conditional probability distribution of departure times, given the arrival time. This correlation between variables is valuable for understanding the utilization pattern of the charging point throughout the day. Additionally, Fig. \ref{fig:evoccur}a indicates that most sessions last no more than 4 hours and are concentrated between 8 am and 6 pm, with some sessions extending into the night. Moreover, Fig. \ref{fig:evoccur}b reveals that more than half of the sessions have a peak power of 0-4 kW, with a charged energy below 10 kWh.
Note that in this analysis of the EV dataset, seasonal impact on charging behaviour is not considered.

\subsection{EV charging session scenarios} \label{EVscenarios}
\begin{figure}[!htbp]
    \centering
    \includegraphics[width=0.97\textwidth]{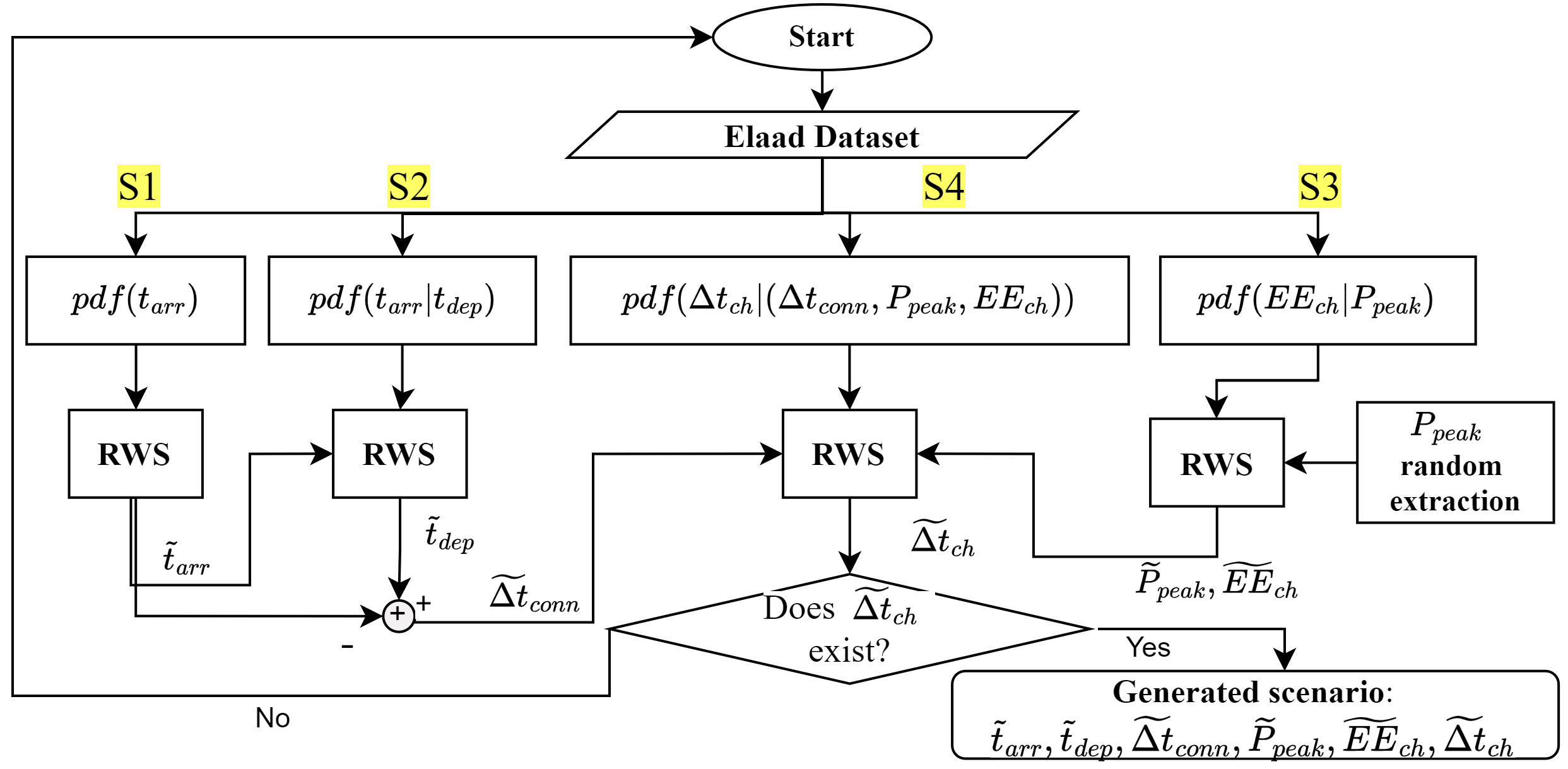}
    \vspace{-6pt}
    \caption{\small{EV charging session scenarios generation.}}
    \label{fig:evscenariosflowchart}
\end{figure}

The flowchart illustrating the generation of EV charging session scenarios is presented in Fig. \ref{fig:evscenariosflowchart}. The arrival time is fixed in S1, next the departure time is fixed based on the conditional probability in S2, the charging energy and time are fixed in S3 and S4 respectively. The approach involves sequentially combining the statistical correlations (steps S1 to S4) identified in Section \ref{EVdataset} with the roulette wheel selection (RWS) \cite{BEHERA2020349} to extract values from the probability distribution functions. Extracted values are denoted by a tilde $\sim$ to distinguish them from random variables.

The flowchart begins with the Elaad dataset to evaluate the necessary probability distribution functions, which are then fed into the RWS steps. The first extraction involves the arrival time, which is subsequently used to extract the departure time through RWS extraction from the corresponding conditional distribution. The difference between arrival and departure time yields the connection time. Peak power is randomly extracted among the considered bins and used in RWS to extract the charged energy from the conditional distribution.

Subsequently, connection time, peak power, and charged energy are employed to extract the charge time from its multi-variable joint conditional distribution. It's worth noting that extracted charge times may not always exist due to no recorded occurrences; in such cases, the scenario is not considered, and the process begins anew.

\begin{figure}[!htbp]
    \centering
    \includegraphics[width=0.9\textwidth]{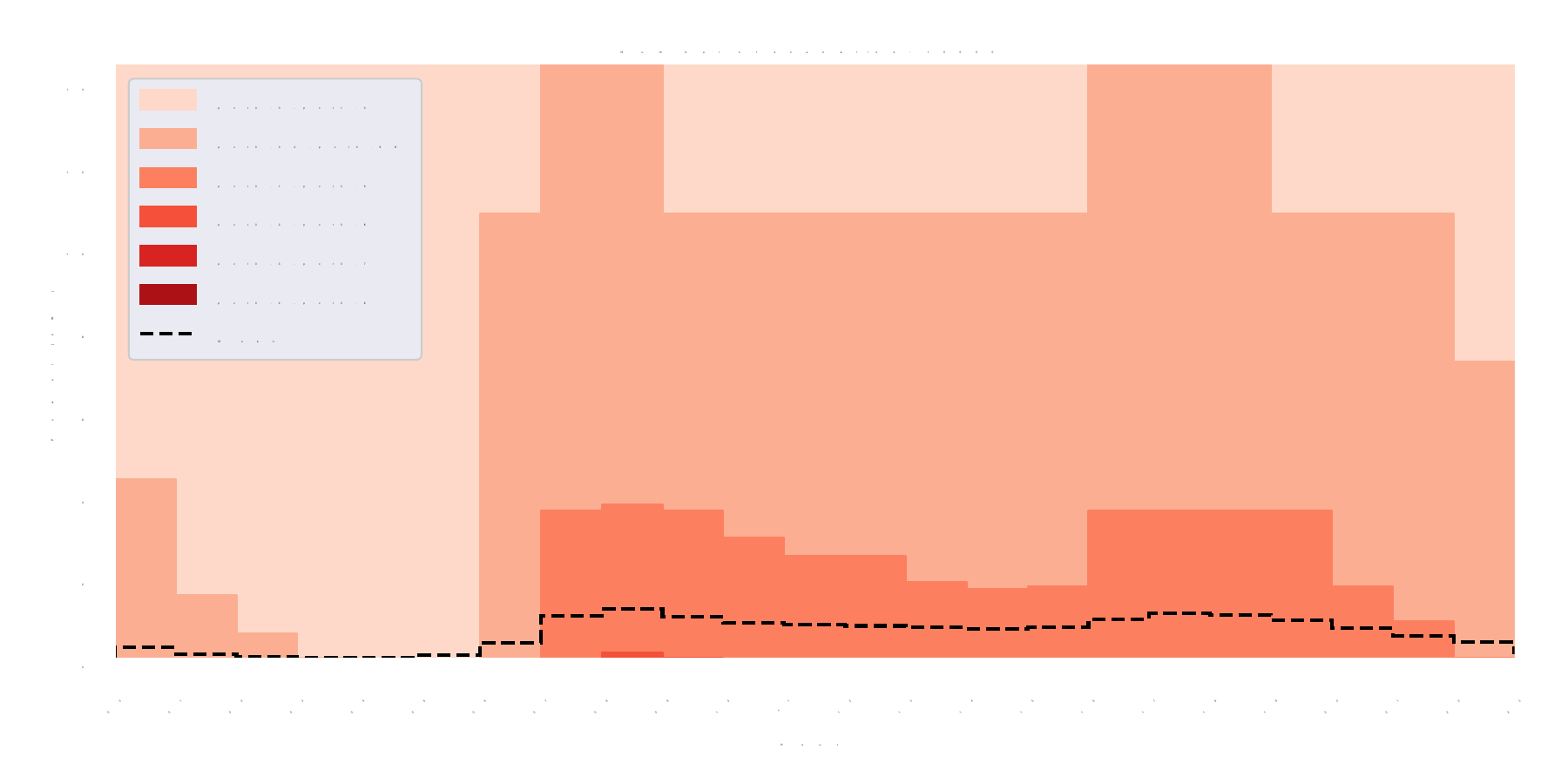}
    \vspace{-10pt}
    \caption{\small{EV charging session scenarios fanchart.}}
    \label{fig:evscenariosfanchart}
\end{figure}

Fig. \ref{fig:evscenariosfanchart} illustrates the significant percentiles of the generated scenarios. Non-smart charging is assumed, where the charging session occurs at nearly maximum power as soon as it begins. The results demonstrate the capability of the developed methodology to replicate the primary features of the Elaad dataset, particularly small peak power and charged energy, minimal activity during the night, and more intense charging sessions for lower percentiles.


\pagebreak

\section{Elia's Solar generation}
\label{section3}
The installed distributed renewable generation is increasing in power networks.
The solar generation in Belgium grew by 1.8 GW in 2023 to 9.9 GW cumulative installed PV \cite{solarPVbelgium}.
Such significant growth necessitates us to assess its impact on the local as well as overall power network. 

In this section, we analyze the publicly available solar generation dataset available by Elia \cite{solarELIA}. The goal is to analyze this dataset for the year 2022 and create quartiles for selecting normalized PV generation profiles. These normalized PV generation profiles are utilized for creating PV generation scenarios based on different levels of installed capacity. This is similar to PV generation scenarios described in \cite{hashmi2023robust}.
The historical data for solar generation provided by Elia includes (a) measured and upscaled PV generation, (b) forecasts at three time ahead levels: (i) hour-ahead, (ii) day-ahead and (iii) week-ahead, (c) monitored installed capacity and (d) the load factor.
The forecast also details P10 and P90 interval forecasts. This additional forecast information can lead to a reduction of risk in decision-making.
\begin{figure}[!htbp]
    \centering
    \includegraphics[width=0.94\textwidth]{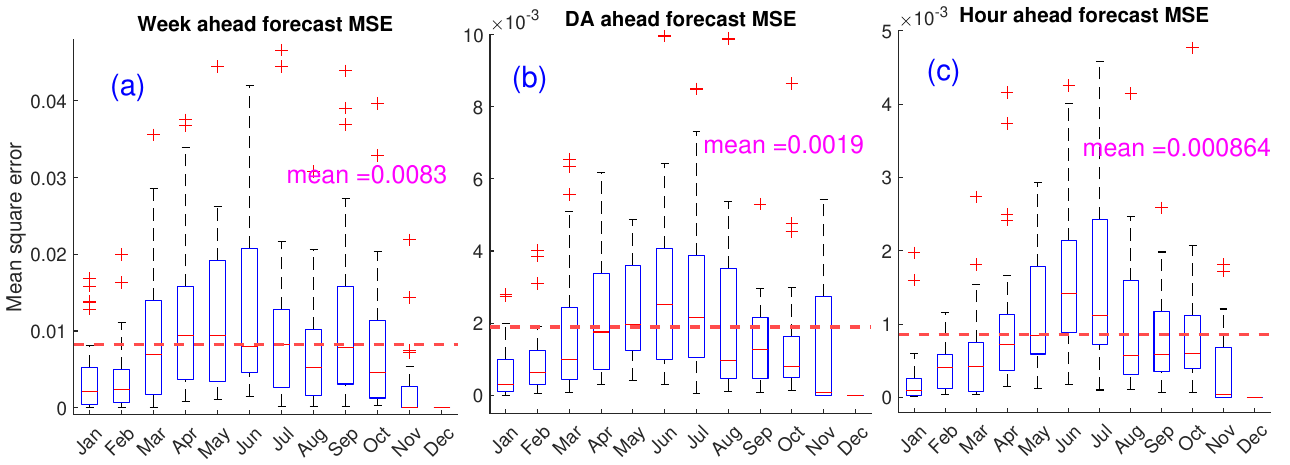}
    \vspace{-4pt}
    \caption{\small{Impact of forecast time and its accuracy: week ahead (WA), day-ahead (DA) and hour ahead (HA).}}
    \label{fig:forecastImpact}
\end{figure}

The distribution of the accuracy of different levels of the time-ahead forecast of solar PV generation is visualized for the year 2022, this is shown in Fig. \ref{fig:forecastImpact}.
It can be observed that the week ahead forecast has a mean square error (MSE) that is 4.4 times that of the DA MSE, and the DA MSE is 2.2 times that of the hour ahead MSE. It shows that forecast accuracy improves with close to real-time solar generation.

\begin{figure}[!htbp]
    \centering
    \includegraphics[width=0.98\textwidth]{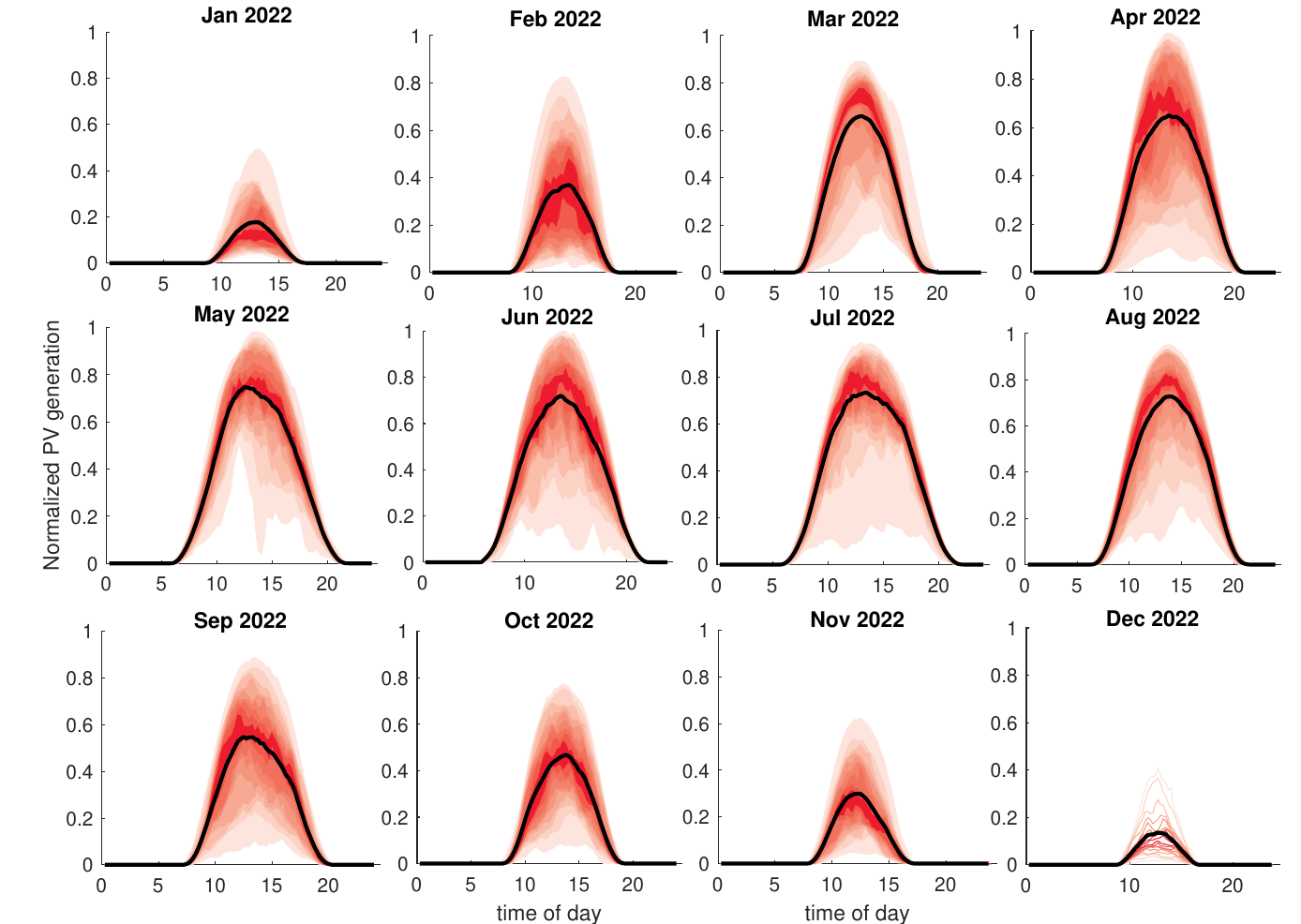}
    \vspace{-4pt}
    \caption{\small{Normalized PV generation in Belgium for 2022.}}
    \label{fig:pvgen}
\end{figure}

PV generation is governed by solar insolation, and the solar insolation fluctuates with seasons. 
Fig. \ref{fig:pvgen} shows the normalized PV generation month-wise for the year 2022.
The monthly PDFs are identified. These PDFs are sampled to identify the seasonal quartiles. These quartiles can be utilized to generate solar generation profiles.
The annual solar generation trends for the year 2022 are shown in Fig. \ref{fig:mcpvgen}. 
Note that these profiles are for Belgium and may vary substantially with geographical variation, this assessment is beyond the scope of this paper.


\pagebreak

\subsection{PV generation scenario creation}
\begin{figure}[!htbp]
    \centering
    \includegraphics[width=0.95\textwidth]{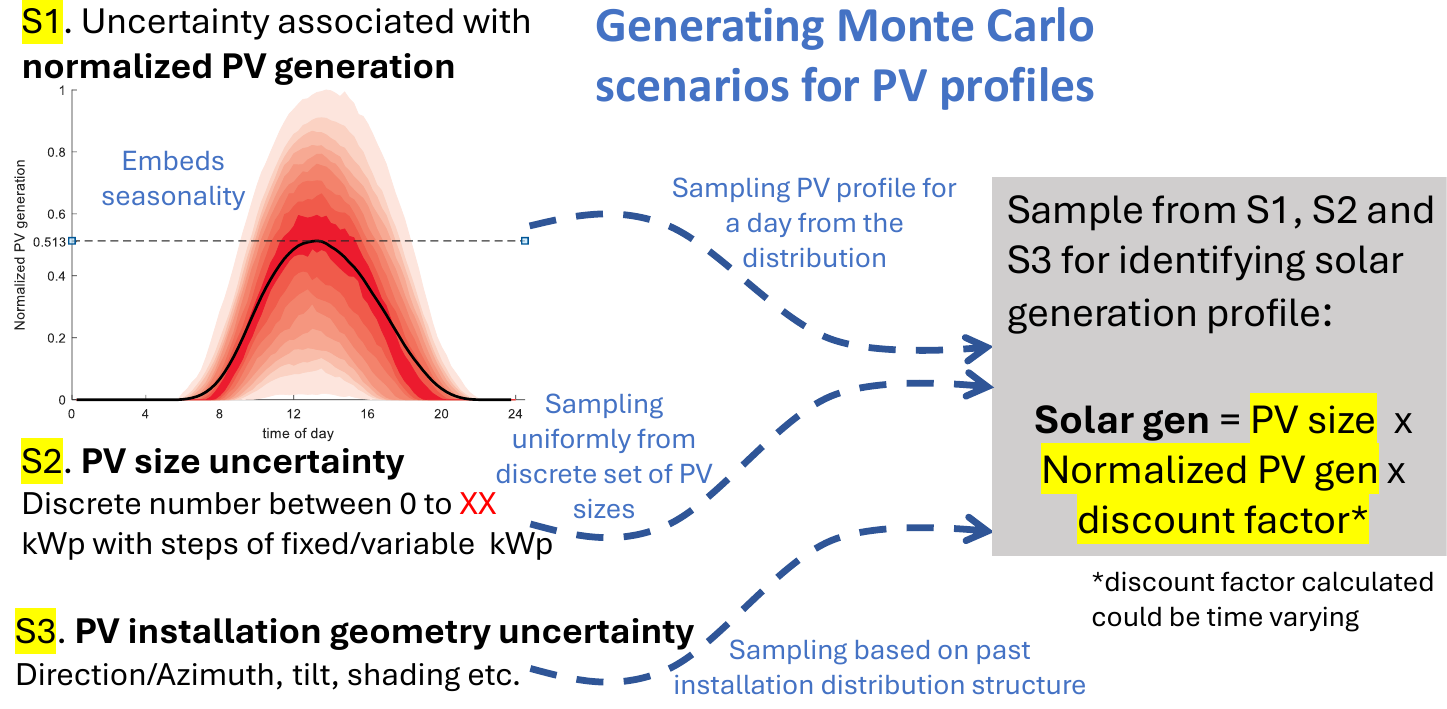}
    \vspace{-4pt}
    \caption{\small{Generating PV generation Monte Carlo scenarios via sampling normalized generation profiles and installed PV size in kWp.}}
    \label{fig:mcpvgen}
\end{figure}

PV generation scenario comprises three decoupled uncertainties: (a) normalized PV profile governed by the solar insolation and temperature, (b) installed kWp, and (c) installed panel's inclination, tilt angle etc.
More DERs are expected to be integrated into the power networks. This proposed scenario generation assists in creating PV generation profiles considering seasonality, temporal fluctuations and stochastic installed PV in kWp. The outline of the PV scenario generation is shown in Fig. \ref{fig:mcpvgen}.
\vspace{-2pt}

\pagebreak


\section{Fluvius load profiles}
\label{section4}
In Feb 2024, Fluvius released a large dataset for 1300 Belgian electricity consumers \cite{FluviusData}.
The Fluvius load profile dataset consists of load profiles for 1300 consumers for the year 2022. This dataset is sampled every 15 minutes and preprocessed to not have any missing data. This high-quality dataset has 5 types of consumers:
\begin{itemize}
\item \textbf{Type 1}: Consumers with no installed EV, PV or HP,
\item \textbf{Type 2}: Consumers with PV installed and no EV or HP,
\item \textbf{Type 3}: Consumers with EV installed and no PV or HP,
\item \textbf{Type 4}: Consumers with PV and HP installed but no EV,
\item \textbf{Type 5}: Consumers with EV and PV installed but no HP.
\end{itemize} 
Type 1 to 5 sets are composed of 300, 300, 100, 300 and 300 consumers, respectively.
The type 1 to 5 categorizing consumers are disjoint set, thus, the total number of consumers with PV, EV and HP are 900, 400 and 300 respectively.
Fluvius consumer type-based metadata is detailed in Tab. \ref{tab:fluvius}.
The metadata identified includes mean max and min net load (NL), mean peak power, mean reverse peak load and metrics for the time and month of the peak and reverse peak load occurrences.

\begin{figure}[!htbp]
    \centering
    \includegraphics[width=0.97\linewidth]{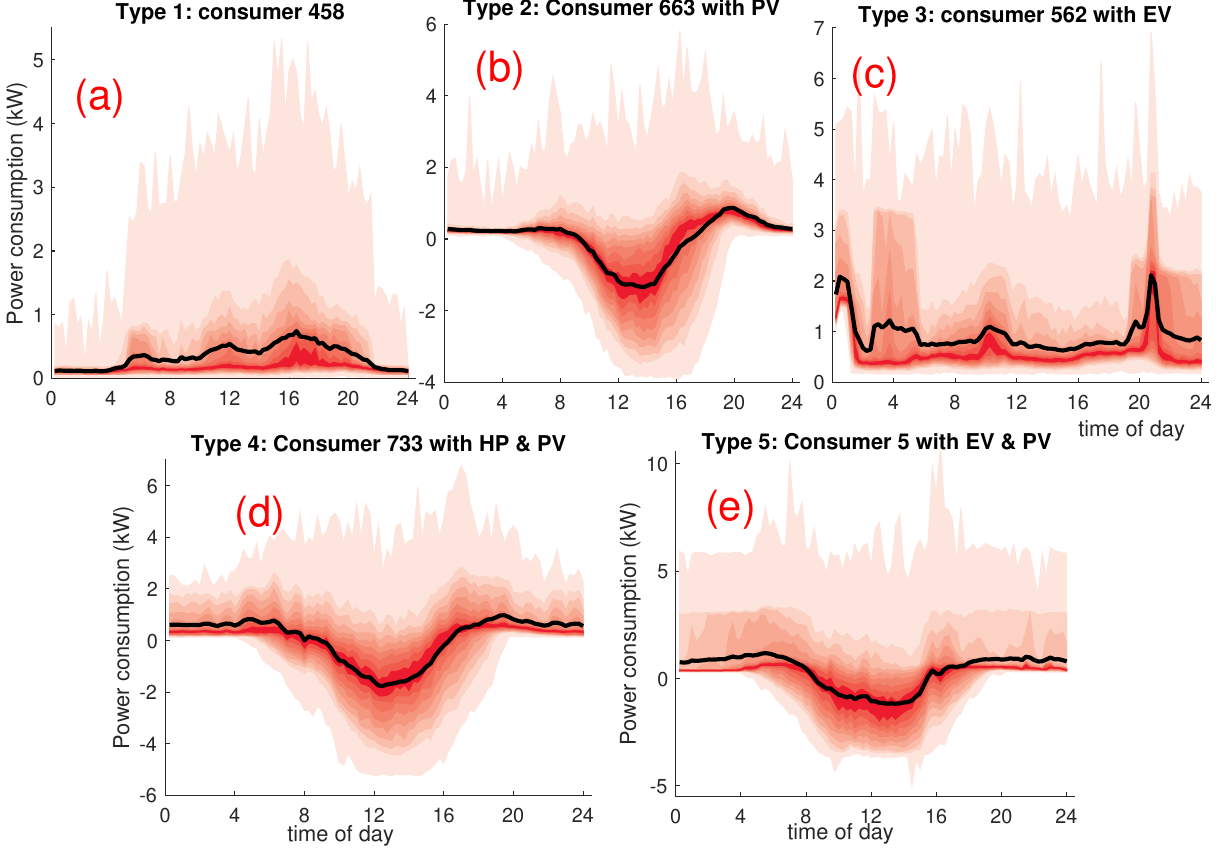}
    \vspace{-4pt}
    \caption{\small{Fluvius representative load profiles for the 5 consumer types}}
    \label{fig:fluviusLoad}
\end{figure}

Fig. \ref{fig:fluviusLoad} shows the representative load profiles for the 5 types of consumers. These consumers are selected based on the distribution of annual consumption kWh.
From Fig. \ref{fig:fluviusLoad}, it is clear that identifying installed solar PV at the consumer end is obvious with the power injection.

\begin{figure}[!htbp]
    \centering
    \includegraphics[width=0.9\textwidth]{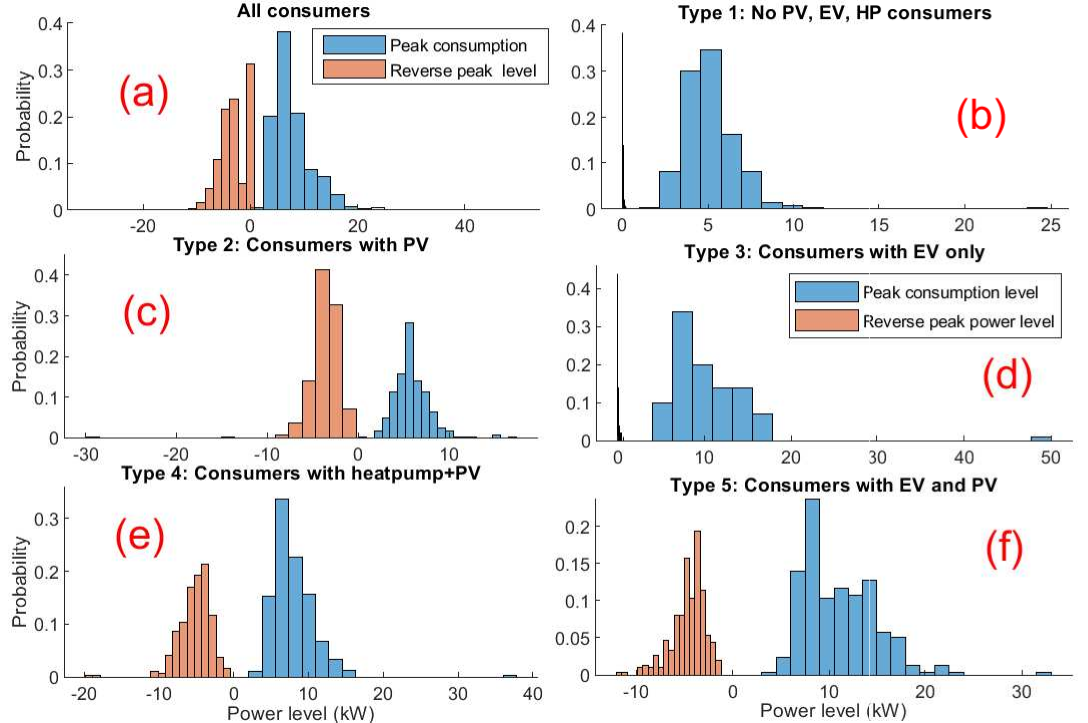}
    \vspace{-4pt}
    \caption{\small{Fluvius load profile (kWp) peaks and reverse peaks}}
    \label{fig:loadpeakFluvius}
\end{figure}


\begin{table*}
\centering
\footnotesize
\caption{{Fluvius data summary}}
\vspace{-2pt}
\label{tab:fluvius}
\begin{tabular}{p{26mm}|p{10mm}|p{6mm}|p{10mm}|p{10mm}|p{10mm}|p{10mm}|p{11mm}|p{10mm}|p{10mm}|p{10mm}|p{10mm}} 
\hline
                                       & \# Consumers & Proba- bility & Mean Net load (kWh) & Max Net kWh & Min~ Net kWh & Mean Peak kW & Mean Reverse kW & Mode Peak time & Peak month & Reverse peak time & Reverse peak month  \\ \hline \hline
Total \# consumers              & 1300            & 1           & 2082.8        & 18091                   & -6025.8                  & 7.8                & -3.1                & 17                  & 12              & 11                & 6                   \\ \hline
\rowcolor{LightCyan}
\textbf{Type 1}: no PV, EV, HP           & 300             & 0.231       & 2943.4        & 9692.2                  & 443.5                    & 5.2                & 0.03                 & 16                  & 1               & 0.8               & 1                   \\ \hline
\rowcolor{LightCyan}
\textbf{Type 2}: only PV                     &      300        &         0.231     &     -278.9           &   5650.8                &   -5695           &         5.93        & -3.7                &      16         &  12               &  12  &        6        \\ \hline
\rowcolor{LightCyan}
\textbf{Type 3}: only EV                     &   100           &  0.231      &   7724.1     &     18091           &         2496.2          &   10.46           &      0.08           &     16.5           &     1          &          8.25       &         6         \\ \hline
\rowcolor{LightCyan}
\textbf{Type 4}: with HP, PV                      & 300             & 0.231       & 765           & 11869.6                 & -6025.8                  & 7.94              & -5.1                & 17                  & 12              & 10.5              & 6                   \\ \hline
\rowcolor{LightCyan}
\textbf{Type 5}: with EV, PV               & 300             & 0.231       & 3021.1        & 13852.8                 & -3894.8                  & 11.3             & -4.6                & 17.3                & 12              & 11                & 6                   \\ \hline
Consumers with PV                      & 900             & 0.692       & 1169.1        & 13852.8                 & -6025.8                  & 8.4              & -4.5                & 17.5                & 12              & 11                & 6                   \\ \hline
Consumers with EV                      & 400             & 0.308       & 4196.9        & 18091                   & -3894.8                  & 11.1             & -3.4                & 17                  & 12              & 11                & 6                   \\ \hline
$\probP$(no EV given PV) & 600             & 0.667       & 243           & 11869.6                 & -6025.8                  & 6.9             & -4.4                & 17                  & 12              & 12                & 6                   \\ \hline
$\probP$(no HP given PV) & 600             & 0.667       & 1371.1        & 13852.8                 & -5694.9                  & 8.6               & -4.1                & 17.5                & 12              & 11                & 6                   \\ \hline
\end{tabular}
\end{table*}

\begin{figure}[!htbp]
    \centering
    \includegraphics[width=0.95\textwidth]{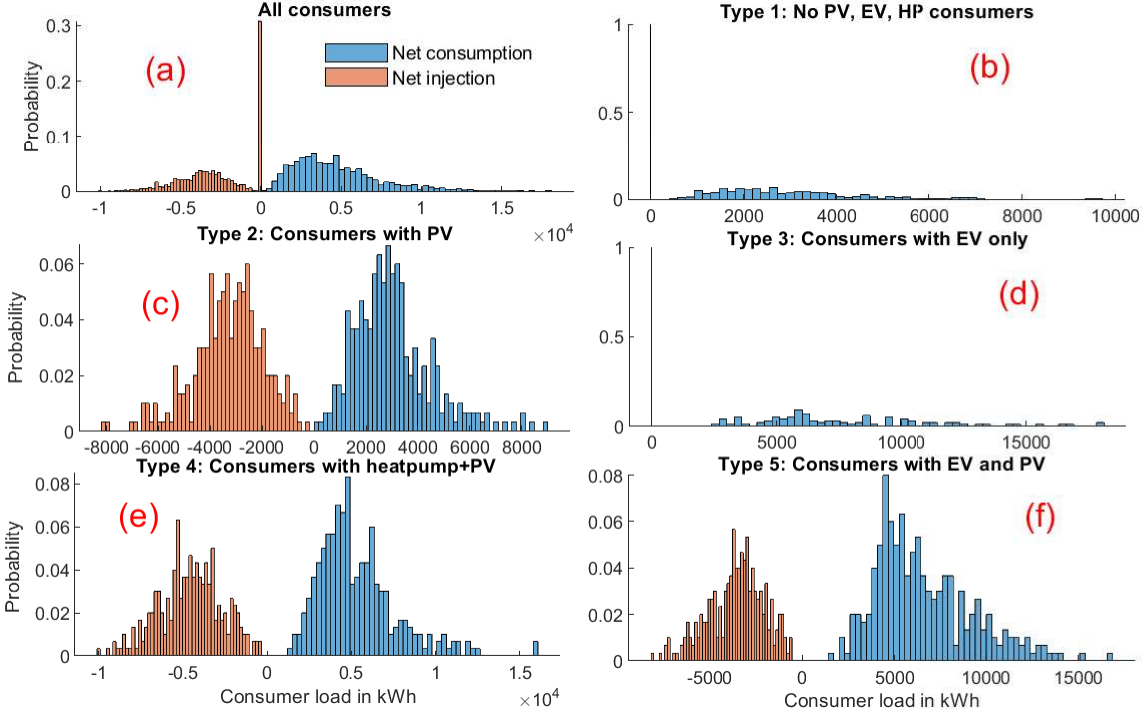}
    \vspace{-4pt}
    \caption{\small{Fluvius load distribution for annual load consumption and reverse consumption (unit of kWh).}}
    \label{fig:consumerdist}
\end{figure}

\begin{figure}[!htbp]
    \centering
    \includegraphics[width=0.95\textwidth]{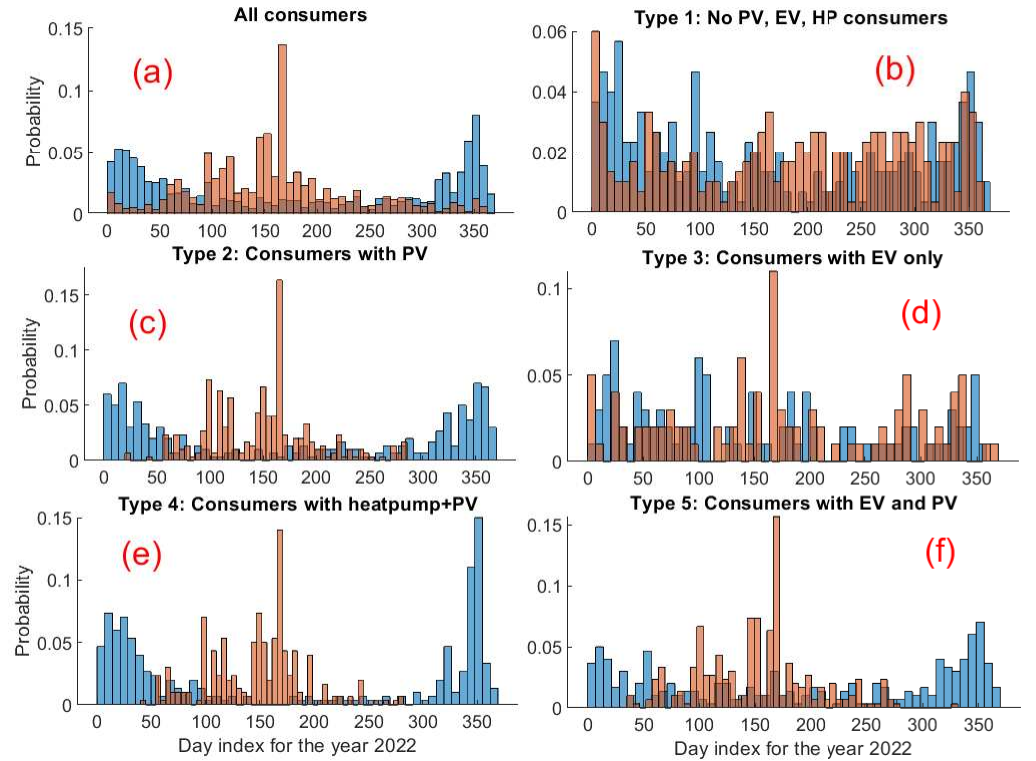}
    \vspace{-4pt}
    \caption{\small{Fluvius load peak and reverse peak (unit of kWp) observed day distribution for 2022.}}
    \label{fig:besandworst}
\end{figure}

\begin{figure*}[!htbp]
    \centering
    \includegraphics[width=0.98\linewidth]{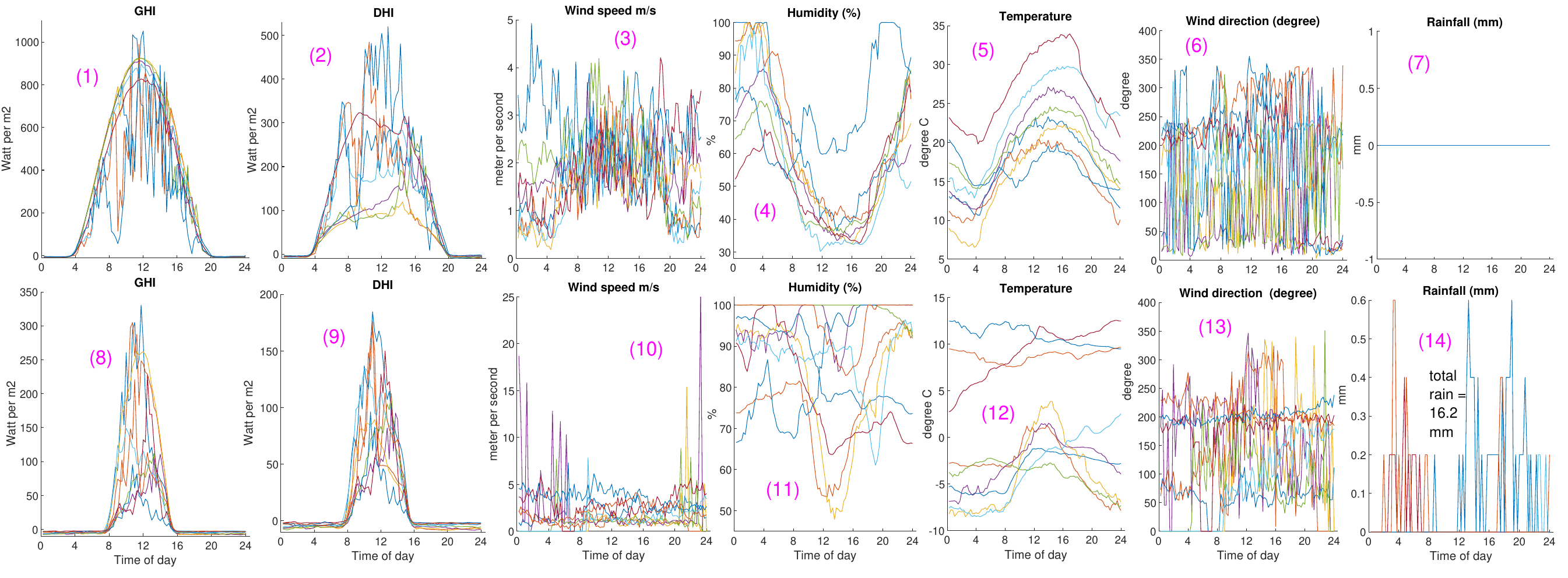}
    \vspace{-8pt}
    \caption{\small{Weather parameter variation for the representative weekdays for days id 163 to 170 (plots 1 to 7) and days 347 to 355 (plots 8 to 14).}}
    \label{fig:weather}
\end{figure*}

\textit{Load peak distribution}:	
For all consumers, around 8\% of the total consumers experienced load peaks between days 347 and 355. For this week, the percentage of different pools of consumers experiencing load peaks are 4.7, 7, 5, 15, and 7\% respectively for types 1 to 5.
Thus, the week with the worst case for the maximum occurrence of load peak instances selected is between days 347 to 355. In this week, around 8\% of consumers had load peaks.

\textit{Reverse load peak distribution}:
For all consumers around 11.5\% of the total consumers experienced reverse peaks between days 166 and 168.
Type 1 consumer reverse peaks are nearly uniformly distributed all throughout the year with the highest number of reverse peaks for the week at the beginning of the year at around 6\%
Type 2 and 5 around 13\% experienced peaks between 166 and 167
Type 3 consumers with only EVs around 10\% experience peaks between 165 and 168.
Type 4 around 14\% experienced reverse load peaks between 166 and 167
Thus, the week with the worst case week for the maximum occurrence of reverse peak instances selected is between days 163 and 170. In this week, around 14\% of consumers had reverse peaks.


\subsection{Weather data for representative weeks}
For the representative weeks 
identified for the Fluvius data is used for filtering the weather data collected at EnergyVille in Belgium. The weather data recorded includes the ambient temperature, wind speed, humidity, wind direction, solar irradiance as GHI and DHI and rainfall.
Further evaluations are needed to quantify the relationships and "causality".
%

\pagebreak












\section{Online repository}
The code developed for datasets analysis, profiles and scenarios generation described in this paper, are publicly available at: 
\url{https://github.com/umar-hashmi/Public-Load-profile-Datasets}.
This repository includes:\\
$\bullet$ \textbf{EV charging dataset}: (i) historical dataset, (ii) probability distribution of charging session characteristics, (iii) scenario generation algorithm, (iv) plots for conditional PDFs.\\
$\bullet$ \textbf{PV generation dataset}: (i) normalized PV profiles: monthly and annual, (ii) Belgium normalized PV generation quartiles.\\
$\bullet$ \textbf{Load profile dataset}: (i) classified consumer types load profile pools, (ii) consumer metadata, (iii) trimming the original dataset from 2.93 GB to 63 MB in size without losing information.\\
$\bullet$ \textbf{Weather dataset} for representative days identified in Sec. \ref{section4}.

\pagebreak

\section{Conclusion and Future work}
\label{section5}
In this paper, we analyzed three publicly available datasets to quantify seasonal and annual uncertainties, facilitating the creation of efficient scenarios. The datasets from Elaad, Elia, and Fluvius are examined to assess electric vehicle charging patterns, normalized solar generation, and low-voltage consumer load profiles, respectively. The load profiles and solar generation data from 2022 are analyzed, incorporating seasonal information. To ensure broader applicability, an online repository has been established. Additionally, the extreme load week(s) are identified and correlated with weather data collected at EnergyVille.

In future work, we will explore efficient scenario generation and scenario reduction techniques for load profiles with EV consumption and PV generation.

\section*{Acknowledgement}
This work is supported by the 
Flemish Government and Flanders Innovation \& Entrepreneurship (VLAIO) through the Flux50 projects InduFlexControl (HBC.2019.0113), and project
IMPROcap (HBC.2022.0733).

We also thank \href{mailto:Georgi.Yordanov@kuleuven.be}{Dr. Georgi Yordanov} for providing the weather measurement dataset at EnergyVille in Genk, Belgium.

\pagebreak

\bibliographystyle{IEEEtran}
\bibliography{reference}

\end{document}